\documentclass[twocolumn,showpacs,prd]{revtex4}
\usepackage{mathrsfs,bm,multirow}
\usepackage{longtable,lscape}
\usepackage{txfonts}
\usepackage{amssymb}
\usepackage{indentfirst}
\usepackage{graphicx,,booktabs}
\usepackage{color}
\usepackage{amssymb}

\usepackage[colorlinks, citecolor=blue,anchorcolor=red,menucolor=red, linkcolor=red,filecolor=red,runcolor=red,urlcolor=blue,frenchlinks=red]{hyperref}

\begin{document}

\title{Prediction of a missing higher charmonium around 4.26 GeV in $J/\psi$ family}
\author{Li-Ping He$^{1,2}$}\email{help08@lzu.edu.cn}
\author{Dian-Yong Chen$^{1,3}$}\email{chendy@impcas.ac.cn}
\author{Xiang Liu$^{1,2}$\footnote{Corresponding author}}\email{xiangliu@lzu.edu.cn}
\author{Takayuki  Matsuki$^{4,5}$}\email{matsuki@tokyo-kasei.ac.jp}
\affiliation{
$^1$Research Center for Hadron and CSR Physics, Lanzhou
University $\&$ Institute of Modern Physics of CAS, Lanzhou 730000,
China\\
$^2$School of Physical Science and Technology, Lanzhou University, Lanzhou 730000, China\\
$^3$Nuclear Theory Group, Institute of Modern Physics of CAS, Lanzhou 730000, China\\
$^4$Tokyo Kasei University, 1-18-1 Kaga, Itabashi, Tokyo 173-8602, Japan\\
$^5$Theoretical Research Division, Nishina Center, RIKEN, Saitama 351-0198, Japan}

\begin{abstract}
Inspired by the similarity between the mass gaps of the $J/\psi$ and $\Upsilon$ families, the prediction of a missing higher charmonium with mass $4263$ MeV and very narrow width is made. In addition, the properties of two charmonium-like states, $X(3940)$ and $X(4160)$, and charmonium $\psi(4415)$ are discussed, where our calculation shows that $X(3940)$ as $\eta_c(3S)$ is established, while the explanation of $X(4160)$ to be $\eta_c(4S)$ is fully excluded and that $\eta_c(4S)$ is typically a very narrow state. These predictions might be accessible at BESIII, Belle, and BelleII in near future.
\end{abstract}


\maketitle

Since the observation of $J/\psi$ in 1974 \cite{Aubert:1974js,Augustin:1974xw}, the charmonium family has become abundant with more and more such states announced by the experiments \cite{Beringer:1900zz}. Especially in the past decade, a series of charmonium-like states
have been observed, which have further stimulated theorists' extensive interest in revealing their underlying properties (see a recent review in Ref. \cite{Liu:2013waa}), since these novel phenomena reflect non-perturbative behavior of quantum chromodynamics (QCD). Among the studies on these states, it is an important research topic for the whole community of particle physics how to identify the exotic states, whose establishment is, of course, tied with our understanding of the charmonium family.

\begin{figure}[htb]
\includegraphics[scale=0.4]{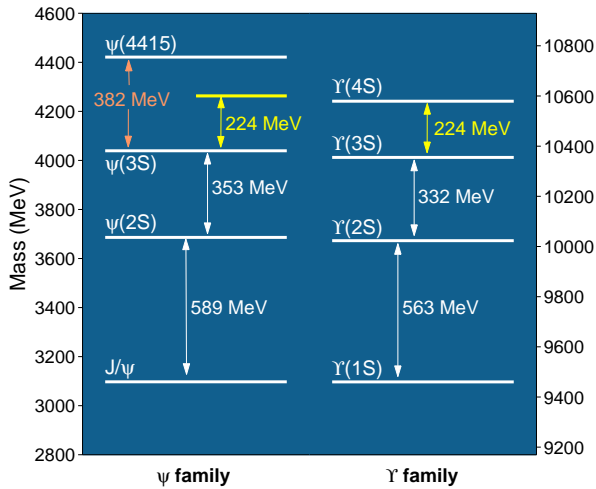}
\caption{(color online). A comparison between the $J/\psi$ and $\Upsilon$ families. \label{mass}}
\end{figure}

When checking the mass spectra of the observed charmonia with spin-parity $J^{PC}=1^{--}$ and comparing them with those of the corresponding bottomonia, we notice that the mass gap between $\psi(2S)$ and $J/\psi$ is almost the same as that between $\Upsilon(2S)$ and $\Upsilon(1S)$. There also exist the similar mass differences, $M_{\psi(3S)}-M_{\psi(2S)}$ and $M_{\Upsilon(3S)}-M_{\Upsilon(2S)}$, where $\psi(2S)$ and $\psi(3S)$ correspond to
$\psi(3686)$ and $\psi(4040)$, respectively.
However, if $\psi(4415)$ is $\psi(4S)$, such a law is violated since the mass gap of $\psi(4415)$ and $\psi(3S)$ is much larger than that of $\Upsilon(4S)$ and $\Upsilon(3S)$. In fact, the properties of the charmonia above 4.1 GeV are still not understood well, which is the possible reason to result in the above puzzling mass gap. In general, compared with the $J/\psi$ family, the bottomonia with the radial quantum numbers $n=1,2,3,4$ were well established both by experiment and theory. Thus, the study of the $J/\psi$ family can be borrowed from $\Upsilon$ family. If this law of mass gap relation still holds for states with $n=3,4$ in $J/\psi$ and $\Upsilon$ families, we find that the mass of $\psi(4S)$ should be located at
$4263$ MeV, where we add the mass gap between $\Upsilon(4S)$ and $\Upsilon(3S)$ to the mass of $\psi(3S)$. In Fig. \ref{mass}, we show the details of the mass gaps for $J/\psi$ and $\Upsilon$ families and compare the corresponding gaps with each other.
After obtaining the above prediction, we notice the results of the mass spectra of charmonium family given in Ref. \cite{Li:2009zu}, where the screening potential was adopted in their calculation. They also found that the mass of $\psi(4S)$ is about 4,273 MeV, which is consistent with our estimate of the mass for the missing charmonium $\psi(4S)$.
Additionally, another theoretical work \cite{Dong:1994zj} also supports existence of a missing $\psi(4S)$. In Ref. \cite{Dong:1994zj}, the color-screening effect was considered in the calculation of the mass spectrum of charmonium, where the mass of $\psi(4S)$ was obtained as 4,247 MeV.

If this predicted state exists in the $J/\psi$ family, we must reveal its underlying properties to answer why there does not have any evidence in the present experiment, which will be the main task of this work.

Before studying this missing charmonium, we need to introduce two charmonium-like states $Y(4260)$ and $Y(4360)$ with the masses around 4,263 MeV. $Y(4260)$ was observed by the BaBar Collaboration in the $J/\psi\pi^+\pi^-$ invariant mass spectrum of $e^+e^-\to J/\psi\pi^+\pi^-$ \cite{Aubert:2005rm}, while $Y(4360)$ was reported by the Belle Collaboration by studying $e^+e^-\to \psi(2S)\pi^+\pi^-$ \cite{Wang:2007ea}. Thus, both of $Y(4260)$ and $Y(4360)$ have $J^{PC}=1^{--}$. We will come back later to discuss whether the missing charmonium has the relation to these two charmonium-like states.

In the following, we study the decay behavior of the missing state corresponding to $\psi(4S)$, which is crucial to explain why this state is still missing at the predicted location and how to search for this in future experiment. In the study we also obtain its full width which is the important information to us.
Since the mass of the missing charmonium predicted (in the next discussion we adopt $\psi(4S)$ to denote this missing charmonium and assume its mass to be 4,263 MeV) is above the thresholds of $D^{(*)}\bar{D}^{(*)}$ and $D_s^{(*)}\bar{D}_s^{(*)}$, $\psi(4S)$ can decay into $D^{(*)}\bar{D}^{(*)}$ and $D_s^{(*)}\bar{D}_s^{(*)}$.

Here, we apply the quark pair creation (QPC) model \cite{Micu:1968mk} to calculate the decay widths, which is one of the most popular models for studying the Okubo-Zweig-Iizuka (OZI) allowed two-body strong decays of hadrons. For the OZI-allowed decay $A\to B+C$, the transition matrix element in terms of the amplitude reads
$\langle
BC|\mathcal{T}|A\rangle=\delta^3(\mathbf{P}_B+\mathbf{P}_C)\mathcal{M}^{M_{J_A}M_{J_B}M_{J_C}}\label{matrix}$
in the center-of-mass frame of the initial state $A$, where $\mathbf{P}_B$ and $\mathbf{P}_C$ are the three-momenta of the final states $B$ and $C$, respectively. In the above expression, $\mathcal{T}$ is the transition operator, i.e.,
\begin{eqnarray} \label{eq2}
\mathcal{T}&=&-3\gamma\sum_m \langle1m; 1-m|00\rangle \int d^3\mathbf{p}_3d^3\mathbf{p}_4\,\delta^3(\mathbf{p}_3+\mathbf{p}_4) \nonumber \\
           &&\times \mathcal{Y}_{1m}\left(\frac{\mathbf{p}_3-\mathbf{p}_4}{2}\right)\,\chi^{34}_{1,-m}\,\phi^{34}_0\,\omega^{34}_0
          \, b^{\dagger}_{3i}(\mathbf{p}_3)\,d^{\dagger}_{4j}(\mathbf{p}_4),
\end{eqnarray}
where $\mathbf{p}_3$ and $\mathbf{p}_4$ denote the momenta carried by the quark and antiquark created from the vacuum.
By using the Jacob-Wick formula \cite{Jacob:1959at}, the partial wave amplitude can be expressed in terms of the amplitude as
\begin{eqnarray}
\mathcal{M}^{JL}(A\rightarrow
BC)&=&\frac{\sqrt{2L+1}}{2J_A+1}\sum_{M_{J_B},M_{J_C}}\langle
L0;JM_{J_A}|J_A M_{J_A}\rangle\nonumber \\ \nonumber
&&\times
\langle J_B M_{J_B};J_C M_{J_C}|J M_{J_A}\rangle
\mathcal{M}^{M_{J_A}M_{J_B}M_{J_C}}(\mathbf{P}).
\end{eqnarray}
Finally the decay width is obtained as
$\Gamma_{A\rightarrow BC}=\pi^2{|\mathbf{P}|}\sum_{J,L}|\mathcal{M}^{JL}|^2/{m_A^2}$.
The above description is just a short summary of the QPC model and the detailed introduction of the QPC model can be found in Refs. \cite{Blundell:1995ev,Blundell:1996as,Luo:2009wu}. We need to specify that in the concrete calculation, the simple harmonic oscillator (SHO) wave function is adopted, where the parameter $R$ is taken from Ref. \cite{Godfrey:1986wj} for final states. Moreover, the dimensionless parameter $\gamma$, describing the strength of the quark-antiquark pair creation, is taken as the same value as that in Ref. \cite{Godfrey:1986wj} by fitting the experimental data.

\begin{figure}[htb]
\includegraphics[scale=0.38]{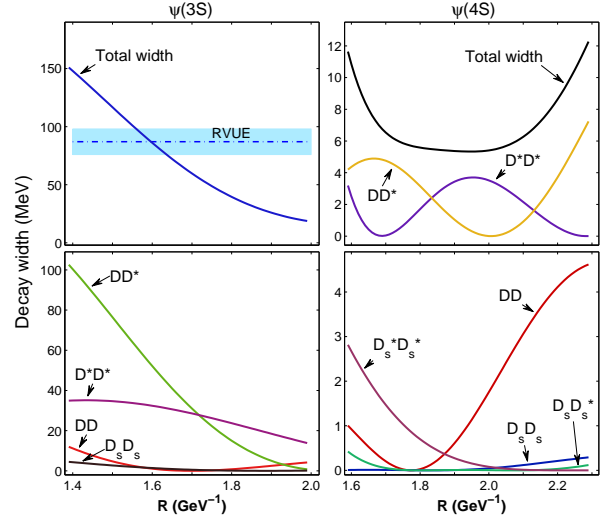}
\caption{(color online). The total and partial decay widths of $\psi(3S)$ (left) and $\psi(4S)$ (right). Here, the dashed line with a band (left) is the experimental data from Ref. \cite{Mo:2010bw}. \label{3S}}
\end{figure}

We first present the result of charmonium $\psi(3S)$ decaying into $D\bar{D}$, $D\bar{D}^*+H.c.$, $D^*\bar{D}^*$ and $D_s\bar{D}_s$ that are kinematically allowed by the phase space of $\psi(3S)$, which is shown in Fig. \ref{3S}. Our calculated result can well reproduce the experimental widths of $\psi(4040)$ in Ref. \cite{Mo:2010bw} when taking $R=1.56-1.63$ GeV$^{-1}$. $D\bar{D}^*+H.c.$ and $D^*\bar{D}^*$ are the dominant decay modes of $\psi(3S)$, while $D\bar{D}$ and $D_s\bar{D}_s$ are the subordinate decays. Here, the conclusion of the $D^{(*)}\bar{D}^{(*)}$ decay modes of $\psi(3S)$ is in agreement with the BaBar data \cite{Aubert:2009aq} since the measured ratio $\Gamma_{D\bar D}/\Gamma_{D^*\bar D+H.c.}$ is $0.24\pm0.05\pm0.12$.
In the same reference, however, the ratio $\Gamma_{D^*\bar{D}^*}/\Gamma_{D^*\bar D+H.c.}$ is given by $0.18\pm0.14\pm0.03$ \cite{Aubert:2009aq} which contradicts with
a larger value of our calculation as shown in Fig. \ref{3S}. Hence this ratio must be tested further in future experiment.
The above study shows that $\psi(4040)$ assigned as $\psi(3S)$ is reasonable. Of course, the reliability of the QPC model is also tested here, which enables us to apply this model to safely study the decays of $\psi(4S)$.

In Fig. \ref{3S}, we give the dependence of the partial decay widths of the predicted $\psi(4S)$ on the $R$ value, which covers the $R$ range discussed in $\psi(3S)$. Here, $D\bar{D}$, $D\bar{D}^*+H.c.$, $D^*\bar{D}^*$, $D_s\bar{D}_s$, $D_s\bar{D}_s^*+H.c.$, $D_s^*\bar{D}_s^*$ are open for $\psi(4S)$. A very interesting result of the decay behavior of $\psi(4S)$ can be found from Fig. \ref{3S}, i.e., the total decay width of $\psi(4S)$ is stable over the corresponding $R$ range adopted, while its partial decay widths strongly depend on the $R$ value. This phenomenon is due to the node effects. Our result also shows that the node effects are important when discussing the higher charmonium decays because due to these effects we find that the predicted charmonium $\psi(4S)$ has very narrow width around 6 MeV.

\begin{figure}[htb]
\includegraphics[scale=0.42]{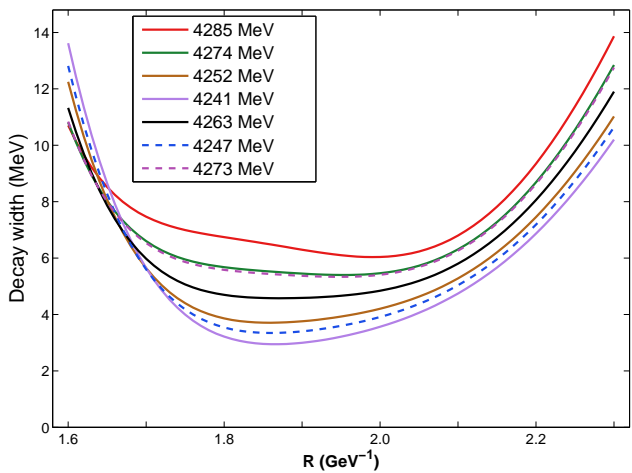}
\caption{(color online). The dependence of the total decay width of $\psi(4S)$ on the mass of $\psi(4S)$. \label{4sde}}
\end{figure}

In addition, we also discuss the dependence of $\psi(4S)$ decay behavior on the mass of $\psi(4S)$ in Fig. \ref{4sde}, where we take several typical values in a range $m_{\psi(4S)}=4,241\sim4,285$ MeV, i.e., $m_{\psi(4S)}=4,241,4,247,4,252,4,273,4,274,4,285$ MeV, which also cover former theoretical predictions of the mass of $\psi(4S)$ in Refs. \cite{Li:2009zu,Dong:1994zj}. The result listed in Fig. \ref{4sde} indicates that the decay behavior of $\psi(4S)$ is weakly dependent on the mass of $\psi(4S)$. Thus, we come to the definite conclusion that $\psi(4S)$ is a narrow state.

For the higher charmonia above the $D\bar{D}$ threshold, this phenomenon of $\psi(4S)$ presented here is unusual since the other higher charmonia, $\psi(4040)$, $\psi(4160)$ and $\psi(4415)$
have widths $80\pm10$ MeV, $103\pm8$ MeV and $62\pm20$ MeV, respectively, all of which are large. Even $\psi(3770)$, which is just 43 MeV above the $D\bar{D}$ threshold, has the width 27.2 MeV.

If considering the large width difference between the predicted $\psi(4S)$ and $Y(4260)/Y(4360)$, it is obvious that there does not exist any correspondence between $\psi(4S)$ and the observed charmonium-like states $Y(4260)/Y(4360)$, where the given average values of widths of $Y(4260)$ and $Y(4360)$ are $95\pm 14$ MeV and $74\pm 18$ MeV in particle data group (PDG) \cite{Beringer:1900zz}, respectively. In Refs. \cite{Chen:2010nv,Chen:2011kc}
the non-resonant explanations to charmonium-like states $Y(4260)$ and $Y(4360)$ were proposed, where both of them can be described by the interference effects of the production amplitudes of $e^+e^-\to J/\psi(\psi(2S))\pi^+\pi^-$ via the intermediate charmonia $\psi(4160)\left(\psi(4415)\right)$ and direct $e^+e^-$ annihilation into $J/\psi(\psi(2S))\pi^+\pi^-$.

As a typical higher charmonium with a very narrow width, the predicted $\psi(4S)$ is difficult to identify by the analysis of the open-charm decay channels \cite{Abe:2006fj,Pakhlova:2008zza,Pakhlova:2007fq,Pakhlova:2009jv} and the $R$ value scan \cite{Burmester:1976mn,Brandelik:1978ei,Siegrist:1981zp,Osterheld:1986hw,Bai:1999pk,Bai:2001ct,CroninHennessy:2008yi,Ablikim:2009ad} based on the present experimental data, which can naturally answer why this higher charmonium is still missing in experiment. Thus, we expect future experimental results of the open-charm decays and more precise study of the $R$ value scan, especially from BESIII, Belle, and forthcoming BelleII.

We notice a recent analysis of BESIII data in Ref. \cite{Yuan:2013ffw}. BESIII already realized the measurement of the cross sections of $e^+e^-\to h_c(1P)\pi^+\pi^-$ at center-of-mass energies $3.90-4.42$ GeV \cite{Ablikim:2013wzq}.  Yuan analyzed the data by fiting the line shape with two Breit-Wigner functions, which indicates that there are a narrow structure with mass $4,216\pm18$ MeV and width $39\pm22$ MeV and another broad structure with mass $4,293\pm9$ MeV and width $222\pm67$ MeV \cite{Yuan:2013ffw}, where these two charmonium-like structures have $J^{PC}=1^{--}$.
Comparing the resonance parameters of this narrow structure in Ref. \cite{Yuan:2013ffw} with our result of the predicted $\psi(4S)$, one finds their similarity, which means that this narrow charmonium-like structure can be as a good candidate of the predicted  $\psi(4S)$
in this work. As mentioned above, more experimental efforts will be necessary to clarify this point.

\begin{figure}
\includegraphics[scale=0.38]{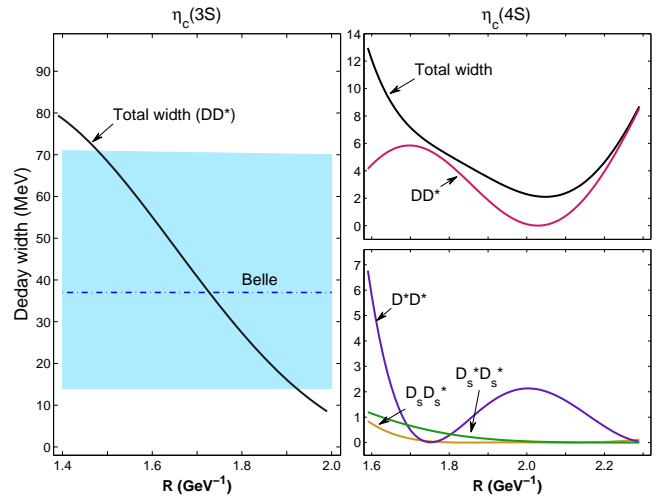}
\caption{(color online). The total and partial decay widths of $\eta_c(3S)$ (left) and $\eta_c(4S)$ (right) and the comparison with the experimental data of charmonium-like state $X(3940)$. Here, the dashed line with a band is the width of $X(3940)$ \cite{Abe:2007sya}. \label{etac}}
\end{figure}

In the following, we apply the above approach to study the $\eta_c$ family by comparing the mass gaps between the $\eta_c$ and $J/\psi$ families, where
the charmonia in $\eta_c$ and $J/\psi$ families belong to S-wave $c\bar{c}$ states. There are two well established $\eta_c$ states listed in PDG, i.e., $\eta_c(1S)$ and $\eta_c(2S)$. Their mass gap is about
$657$ MeV, which is similar to that between $\psi(2S)$ and $\psi(1S)$. Thus, assuming the mass gap between $\eta_c(3S)$ and $\eta_c(2S)$ is the same as that between $\psi(3S)$ and $\psi(2S)$, we can estimate the mass of $\eta_c(3S)$ as about 3,992 MeV, which is close to the mass of an observed charmonium-like state $X(3940)$ in the $D^*\bar{D}$ invariant mass spectrum of $e^+e^-\to D^*\bar{D} J/\psi$ \cite{Abe:2007jna,Abe:2007sya}. Here, the $C$ parity of $X(3940)$ favors $+1$ since $X(3940)$ is from the double charm production. $\eta_c$ also has the same $C$ parity $+1$. Thus, we may identify $X(3940)$ as $\eta_c(3S)$. To test it, we also study the decay behavior of $X(3940)$ under the $\eta_c(3S)$ assignment, which is listed in Fig. \ref{etac}. We find that the total decay width of $\eta_c(3S)$ calculated can be well fitted with the experimental value of $X(3940)$ when we take the same $R$ range as that in the study of $\psi(3S)$, where $D\bar{D}^*+H.c.$ is the only decay mode of $\eta_c(3S)$, which can explain why $X(3940)$ was observed in the $D\bar{D}^*$ channel \cite{Abe:2007jna,Abe:2007sya} under this assumption.

In addition, another charmonium-like state $X(4160)$ was reported by the Belle Collaboration through the double charm production, where $X(4160)$ appears in the $D^*\bar{D}^*$ invariant mass distribution of $e^+e^-\to D^{*+}{D}^{*-}J/\psi$ \cite{Abe:2007sya}. In this work, we also discuss whether $X(4160)$ can be explained as $\eta_c(4S)$. In Fig. \ref{etac}, the partial and total decay widths of $\eta_c(4S)$ are listed. The situation of the decay behavior of $\eta_c(4S)$ is very similar to that of $\psi(4S)$ discussed above. Our study shows that the total width of $\eta_c(4S)$ is also very narrow, which is not consistent with the measured width of $X(4160)$ that has $139^{+111}_{-61}\pm21$ MeV \cite{Abe:2007sya}. According to the above analysis, we can fully exclude the $\eta_c(4S)$ assignment to $X(4160)$.

Before summarizing our work, we would like to pay attention to $\psi(4415)$. Here, we discuss the possibility of $\psi(4415)$ as $\psi(5S)$.

\begin{figure}
\includegraphics[scale=0.38]{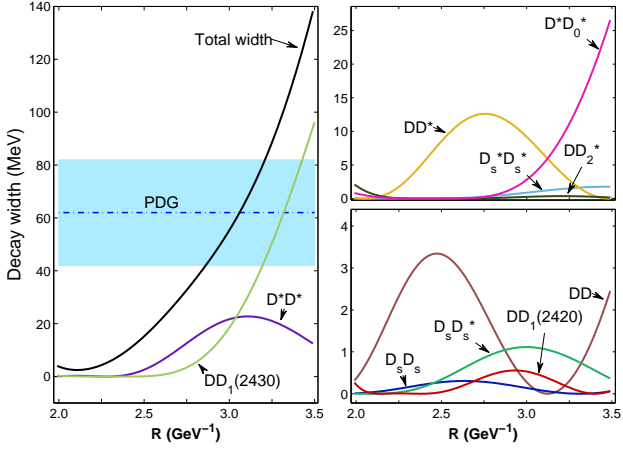}
\caption{(color online). The total and two-body strong decays of $\psi(4415)$ as $\psi(5S)$. Here, the dashed line with a band is the experimental data taken from PDG \cite{Beringer:1900zz}.  \label{5S}}
\end{figure}

The two-body strong decay behaviors of $\psi(4415)$ as a $\psi(5S)$ state are presented in Fig. \ref{5S}. The total width overlaps with the average width in PDG when $R$ is in the range of 2.86 to 3.21 GeV$^{-1}$. $D^*\bar{D}^*$, $D\bar{D}_1(2430)+H.c.$ and $D\bar{D}_1(2420)+H.c.$ are its main decay channels. Additionally, the typical ratios are $R_1=\Gamma_{D\bar D}/\Gamma_{D^*\bar D^*}=0.004-0.06$ and $R_2=\Gamma_{D^*\bar D+H.c.}/\Gamma_{D^*\bar D^*}=0.19-0.68$ corresponding to $R=2.86-3.19$ GeV$^{-1}$, which well agree with the BaBar data \cite{Aubert:2009aq} having the measured values of $R_1=0.14\pm0.12\pm0.03$ and $R_2=0.17\pm0.25\pm0.03$ \cite{Aubert:2009aq}. The above analysis indicates that it is very likely that $\psi(4415)$ is the $\psi(5S)$ state. However, in our calculation the branching ratio $\Gamma_{D\bar D_2^*}/\Gamma_{total}$ is about two orders of magnitude smaller than the experimental data from Belle \cite{Pakhlova:2007fq}. This incompatibility is a challenge to the explanation of $\psi(4415)$ as $\psi(5S)$ state. We also suggest the experimental study of $\psi(4415)$ in future experiments, which is important to further understand the properties of $\psi(4415)$.

For the convenience of reader,  in Table \ref{Tab:theo} we further collect the experimental information of these charmonium-like states or charmonium involved in the present work.

\renewcommand{\arraystretch}{1.6}
\begin{table}[htb]
\centering
\caption{The experimental information of $X(3940)$, $X(4160)$, $Y(4260)$, $Y(4360)$, $\psi(4415)$ and
and possible $Y(4216)$.
\label{Tab:theo}}
\begin{tabular}{cccccc}
\toprule[1pt]
States&Mass (MeV)&Width (MeV)&Observed channel\\\midrule[1pt]
$X(3940)$ \cite{Abe:2007jna,Abe:2007sya}&$3942^{+7}_{-6}\pm6$ &$37^{+26}_{-15}\pm8$&$e^+e^-\to D^*\bar{D}J/\psi$\\
$X(4160)$ \cite{Abe:2007sya}&$4156^{+25}_{-20}\pm15$ &$139^{+111}_{-61}\pm21$&$e^+e^-\to D^{*+}{D}^{*-}J/\psi$\\
$Y(4260)$ \cite{Aubert:2005rm}&$4259\pm8^{+2}_{-6}$&$88\pm23^{+6}_{-4}$&$e^+e^-\to J/\psi\pi^+\pi^-$\\
$Y(4360)$ \cite{Wang:2007ea}&$4361\pm 9\pm 9$&$74\pm 15\pm 10$&$e^+e^-\to \psi(2S)\pi^+\pi^-$\\
$\psi(4415)$ \cite{Beringer:1900zz}&$4421\pm4$&$62\pm20$&see PDG \cite{Beringer:1900zz} for details\\
$Y(4216)$ \cite{Ablikim:2013wzq,Yuan:2013ffw}&$4216\pm18$&$39\pm22$&$e^+e^-\to h_c\pi^+\pi^-$\\
\bottomrule[1pt]
\end{tabular}
\end{table}

In summary, in this work we have predicted a missing charmonium $\psi(4S)$ with quantum number $n^{2S+1}L_J=4^3 S_1$, which has mass around $4263$ MeV and very narrow width. This observation has been obtained through the similarity of the mass gaps existing in $J/\psi$ and $\Upsilon$ families and further study of its OZI-allowed decay behavior. Comparing this state with the reported higher charmonia $\psi(4040)$, $\psi(4160)$ and $\psi(4415)$, the predicted $\psi(4S)$ state is the first higher charmonium with such a narrow width, which can explain why there is no evidence in the corresponding analysis of the open-charm decay channels and the $R$ value scan until now. The relation of $\psi(4S)$ and the recent evidence of a narrow charmonium-like structure in the $e^+e^-\to h_c(1P)\pi^+\pi^-$ process \cite{Yuan:2013ffw} has also been discussed.

We have also explored two charmonium-like states $X(3940)$ and $X(4160)$ under the $\eta_c(3S)$ and $\eta_c(4S)$ assignments, respectively. Our study has indicated that $X(3940)$ can be well explained as $\eta_c(3S)$ but $X(4160)$ as $\eta_c(4S)$ is fully excluded. Here, we have predicted $\eta_c(4S)$ also has very narrow width similar to the situation of the discussed $\psi(4S)$. In addition, the properties of $\psi(4415)$ as $\psi(5S)$ have been given in this work.

The study presented in this work can enhance our understanding of charmonium family especially for $J/\psi$ family, which is valuable to reveal the underlying QCD non-perturbative effects.  To test our predictions, we expect further experimental progress on higher charmonia, where BESIII, Belle, and BelleII will be good platforms to
carry out the search for our predictions.

\vfil

\section*{Acknowledgements}

This project is supported by the National Natural Science Foundation of China under Grants No. 11222547, No. 11175073, No. 11035006 and No. 11375240, the Ministry of Education of China (FANEDD under Grant No. 200924, SRFDP under Grant No. 20120211110002, NCET, the Fundamental Research Funds for the Central Universities), the Fok Ying Tung Education Foundation (No. 131006). X.L. would like to expresses his sincere thanks to Professor Takayuki Matsuki for providing the agreeable atmosphere during his stay at Tokyo Kasei University.

\vfil


\begin{thebibliography}{99}
\bibitem{Aubert:1974js}
  J.~J.~Aubert {\it et al.}  [E598 Collaboration],
  Phys.\ Rev.\ Lett.\  {\bf 33}, 1404 (1974).

\bibitem{Augustin:1974xw}
  J.~E.~Augustin {\it et al.}  [SLAC-SP-017 Collaboration],
  Phys.\ Rev.\ Lett.\  {\bf 33}, 1406 (1974).

\bibitem{Beringer:1900zz}
  K.~A.~Olive {\it et al.}  [Particle Data Group Collaboration],
  Chin.\ Rev.\ C {\bf 38}, 090001 (2014).

\bibitem{Liu:2013waa}
  X.~Liu,
  arXiv:1312.7408 [hep-ph].

\bibitem{Li:2009zu}
  B.~-Q.~Li and K.~-T.~Chao,
  Phys.\ Rev.\ D {\bf 79}, 094004 (2009)
  [arXiv:0903.5506 [hep-ph]].

\bibitem{Dong:1994zj}
  Y.~B.~Dong, Y.~W.~Yu, Z.~Y.~Zhang and P.~N.~Shen,
  Phys.\ Rev.\ D {\bf 49}, 1642 (1994).

\bibitem{Aubert:2005rm}
  B.~Aubert {\it et al.}  [BaBar Collaboration],
  Phys.\ Rev.\ Lett.\  {\bf 95}, 142001 (2005)
  [hep-ex/0506081].

\bibitem{Wang:2007ea}
  X.~L.~Wang {\it et al.}  [Belle Collaboration],
  Phys.\ Rev.\ Lett.\  {\bf 99}, 142002 (2007)
  [arXiv:0707.3699 [hep-ex]].

\bibitem{Micu:1968mk}
  L.~Micu,
  Nucl.\ Phys.\ B {\bf 10}, 521 (1969).

\bibitem{Jacob:1959at}
  M.~Jacob and G.~C.~Wick,
  Annals Phys.\  {\bf 7}, 404 (1959)  [Annals Phys.\  {\bf 281}, 774 (2000)].  

\bibitem{Blundell:1995ev}
  H.~G.~Blundell and S.~Godfrey,
  Phys.\ Rev.\ D {\bf 53}, 3700 (1996)
  [hep-ph/9508264].

\bibitem{Blundell:1996as}
  H.~G.~Blundell,
  hep-ph/9608473.

\bibitem{Luo:2009wu}
  Z.~-G.~Luo, X.~-L.~Chen and X.~Liu,
  Phys.\ Rev.\ D {\bf 79}, 074020 (2009)
  [arXiv:0901.0505 [hep-ph]].

\bibitem{Godfrey:1986wj}
  S.~Godfrey and R.~Kokoski,
  Phys.\ Rev.\ D {\bf 43}, 1679 (1991).

\bibitem{Mo:2010bw}
  X.~H.~Mo, C.~Z.~Yuan and P.~Wang,
  Phys.\ Rev.\ D {\bf 82}, 077501 (2010)
  [arXiv:1007.0084 [hep-ex]].

\bibitem{Aubert:2009aq}
  B.~Aubert {\it et al.}  [BaBar Collaboration],
  Phys.\ Rev.\ D {\bf 79}, 092001 (2009)
  [arXiv:0903.1597 [hep-ex]].

\bibitem{Chen:2010nv}
  D.~Y.~Chen, J.~He and X.~Liu,
  Phys.\ Rev.\ D {\bf 83}, 054021 (2011)
  [arXiv:1012.5362 [hep-ph]].

\bibitem{Chen:2011kc}
  D.~Y.~Chen, J.~He and X.~Liu,
  Phys.\ Rev.\ D {\bf 83}, 074012 (2011)
  [arXiv:1101.2474 [hep-ph]].


\bibitem{Abe:2006fj}
  K.~Abe {\it et al.}  [Belle Collaboration],
  Phys.\ Rev.\ Lett.\  {\bf 98}, 092001 (2007)  [hep-ex/0608018].  

\bibitem{Pakhlova:2008zza}
  G.~Pakhlova {\it et al.}  [Belle Collaboration],
  Phys.\ Rev.\ D {\bf 77}, 011103 (2008)  [arXiv:0708.0082 [hep-ex]].  

\bibitem{Pakhlova:2007fq}
  G.~Pakhlova {\it et al.}  [Belle Collaboration],
  Phys.\ Rev.\ Lett.\  {\bf 100}, 062001 (2008)  [arXiv:0708.3313 [hep-ex]].  

\bibitem{Pakhlova:2009jv}
  G.~Pakhlova {\it et al.}  [Belle Collaboration],
  Phys.\ Rev.\ D {\bf 80}, 091101 (2009)  [arXiv:0908.0231 [hep-ex]].  

\bibitem{Burmester:1976mn}
  J.~Burmester {\it et al.}  [PLUTO Collaboration],
  Phys.\ Lett.\ B {\bf 66}, 395 (1977).  

\bibitem{Brandelik:1978ei}
  R.~Brandelik {\it et al.}  [DASP Collaboration],
  Phys.\ Lett.\ B {\bf 76}, 361 (1978).

\bibitem{Siegrist:1981zp}
  J.~Siegrist, R.~Schwitters, M.~S.~Alam, A.~Boyarski, M.~Breidenbach, F.~Bulos, J.~T.~Dakin and J.~Dorfan {\it et al.},
  Phys.\ Rev.\ D {\bf 26}, 969 (1982).  

\bibitem{Osterheld:1986hw}
  A.~Osterheld, R.~Hofstadter, R.~Horisberger, I.~Kirkbride, H.~Kolanoski, K.~Konigsmann, A.~Liberman and J.~O'Reilly {\it et al.},
  SLAC Report No. SLAC-PUB-4160,1986

\bibitem{Bai:1999pk}
  J.~Z.~Bai {\it et al.}  [BES Collaboration],
  Phys.\ Rev.\ Lett.\  {\bf 84}, 594 (2000)  [hep-ex/9908046].  

\bibitem{Bai:2001ct}
  J.~Z.~Bai {\it et al.}  [BES Collaboration],
  Phys.\ Rev.\ Lett.\  {\bf 88}, 101802 (2002)  [hep-ex/0102003].  

\bibitem{CroninHennessy:2008yi}
  D.~Cronin-Hennessy {\it et al.}  [CLEO Collaboration],
  Phys.\ Rev.\ D {\bf 80}, 072001 (2009)  [arXiv:0801.3418 [hep-ex]].  

\bibitem{Ablikim:2009ad}
  M.~Ablikim {\it et al.}  [BES Collaboration],
  Phys.\ Lett.\ B {\bf 677}, 239 (2009)  [arXiv:0903.0900 [hep-ex]].  







\bibitem{Yuan:2013ffw}
  C.~-Z.~Yuan,
  Chin.\ Phys.\ C {\bf 38}, 043001 (2014)
  [arXiv:1312.6399 [hep-ex]].

\bibitem{Ablikim:2013wzq}
  M.~Ablikim {\it et al.}  [BESIII Collaboration],
  Phys.\ Rev.\ Lett.\  {\bf 111}, 242001 (2013)
  [arXiv:1309.1896 [hep-ex]].



\bibitem{Abe:2007jna}
  K.~Abe {\it et al.}  [Belle Collaboration],
  Phys.\ Rev.\ Lett.\  {\bf 98}, 082001 (2007)
  [hep-ex/0507019].

\bibitem{Abe:2007sya}
  P.~Pakhlov {\it et al.}  [Belle Collaboration],
  Phys.\ Rev.\ Lett.\  {\bf 100}, 202001 (2008)
  [arXiv:0708.3812 [hep-ex]].


\end{thebibliography}
\end{document}